\documentclass[sigconf]{acmart}
\usepackage{amsthm}
\usepackage{subcaption}
\usepackage{paralist}
\usepackage{bm}
\usepackage{amsfonts}
\usepackage{multirow}
\usepackage{cases}
\usepackage{booktabs}
\usepackage{threeparttable}
\usepackage{epstopdf}
\usepackage{upgreek}
\usepackage{endnotes}
\usepackage{etoolbox}
\usepackage{algpseudocode}
\usepackage{amsmath}
\usepackage[ruled]{algorithm2e}
\usepackage{balance}
\usepackage{xspace}
\usepackage{array}
\usepackage{enumitem}
\usepackage{hyperref}
\AtBeginDocument{%
  \providecommand\BibTeX{{%
    \normalfont B\kern-0.5em{\scshape i\kern-0.25em b}\kern-0.8em\TeX}}}
\newcommand{\eat}[1]{}
\newcommand{\paratitle}[1]{\vspace{1.5ex}\noindent \textbf{#1}}

\newcommand{\baby}{\textsc{Maria}\xspace}

\copyrightyear{2023} 
\acmYear{2023} 
\setcopyright{acmlicensed}\acmConference[SIGIR '23]{Proceedings of the 46th International ACM SIGIR Conference on Research and Development in Information Retrieval}{July 23--27, 2023}{Taipei, Taiwan}
\acmBooktitle{Proceedings of the 46th International ACM SIGIR Conference on Research and Development in Information Retrieval (SIGIR '23), July 23--27, 2023, Taipei, Taiwan}
\acmPrice{15.00}
\acmDOI{10.1145/3539618.3591736}
\acmISBN{978-1-4503-9408-6/23/07}

\begin{document}
\title{Multi-Scenario Ranking with Adaptive Feature Learning}

\author{Yu Tian}
\email{s.braylon1002@gmail.com}
\affiliation{%
  \institution{Key Laboratory of Aerospace Information Security and Trusted Computing, Ministry of Education, School of Cyber
Science and Engineering, Wuhan University}
  \country{China}
}

\author{Bofang Li}
\email{bofang.lbf@alibaba-inc.com}
\affiliation{%
  \institution{Alibaba Group}
  \country{China}
}

\author{Si Chen}
\email{yasui.cs@alibaba-inc.com}
\affiliation{%
  \institution{Alibaba Group}
  \country{China}
}

\author{Xubin Li}
\email{lxb204722@alibaba-inc.com}
\affiliation{%
  \institution{Alibaba Group}
  \country{China}
}

\author{Hongbo Deng}
\email{dhb167148@alibaba-inc.com}
\affiliation{%
  \institution{Alibaba Group}
  \country{China}
}

\author{Jian Xu}
\email{xiyu.xj@alibaba-inc.com}
\affiliation{%
  \institution{Alibaba Group}
  \country{China}
}

\author{Bo Zheng}
\email{bozheng@alibaba-inc.com}
\affiliation{%
  \institution{Alibaba Group}
  \country{China}
}

\author{Qian Wang}
\email{	qianwang@whu.edu.cn}
\affiliation{%
  \institution{Key Laboratory of Aerospace Information Security and Trusted Computing, Ministry of Education, School of Cyber
Science and Engineering, Wuhan University}
  \country{China}
}

\author{Chenliang Li}\authornote{Corresponding author. Work done when Yu Tian was an intern at Alibaba.}
\email{cllee@whu.edu.cn}
\affiliation{%
  \institution{Key Laboratory of Aerospace Information Security and Trusted Computing, Ministry of Education, School of Cyber
Science and Engineering, Wuhan University}
  \country{China}
}
\renewcommand{\shortauthors}{Yu Tian et al.}

\begin{abstract}
 Recently, Multi-Scenario Learning (MSL) is widely used in recommendation and retrieval systems in the industry because it facilitates transfer learning from different scenarios, mitigating data sparsity and reducing maintenance cost. These efforts produce different MSL paradigms by searching more optimal network structure, such as Auxiliary Network, Expert Network, and Multi-Tower Network. It is intuitive that different scenarios could hold their specific characteristics, activating the user's intents quite differently. In other words, different kinds of auxiliary features would bear varying importance under different scenarios. With more discriminative feature representations refined in a scenario-aware manner, better ranking performance could be easily obtained without expensive search for the optimal network structure. Unfortunately, this simple idea is mainly overlooked but much desired in real-world systems.
 
 To this end, in this paper, we propose a \textbf{m}ulti-scen\textbf{a}rio \textbf{r}anking framework with adapt\textbf{i}ve fe\textbf{a}ture learning (named \baby). Specifically, \baby is devised to inject the scenario semantics in the bottom part of the network to derive more discriminative feature representations. There are three components designed in \baby for this purpose: \textit{feature scaling}, \textit{feature refinement}, and \textit{feature correlation modeling}. The purpose of feature scaling is to highlight the scenario-relevant fields and also suppress the irrelevant ones. Then, the feature refinement utilizes an automatic refiner selection subnetwork for each feature field, such that the high-level semantics with respect to the scenario can be extracted with the optimal expert. Afterwards, we further explicitly derive the feature correlations across fields as complementary signals. The resultant representations are then fed into a simple MoE structure with an additional scenario-shared tower for final prediction. Experiments on two large-scale real-world datasets demonstrate the superiority of \baby against several state-of-the-art baselines for both product search and recommendation. Further analysis also validates the rationality of adaptive feature learning under a multi-scenario scheme. Moreover, our A/B test results on the Alibaba search advertising platform also demonstrate that \baby is superior in production environments.
\end{abstract}



\maketitle
\section{Introduction}\label{sec:intro}

\begin{figure}[!]
    \centering
    \includegraphics[width=0.95\linewidth]{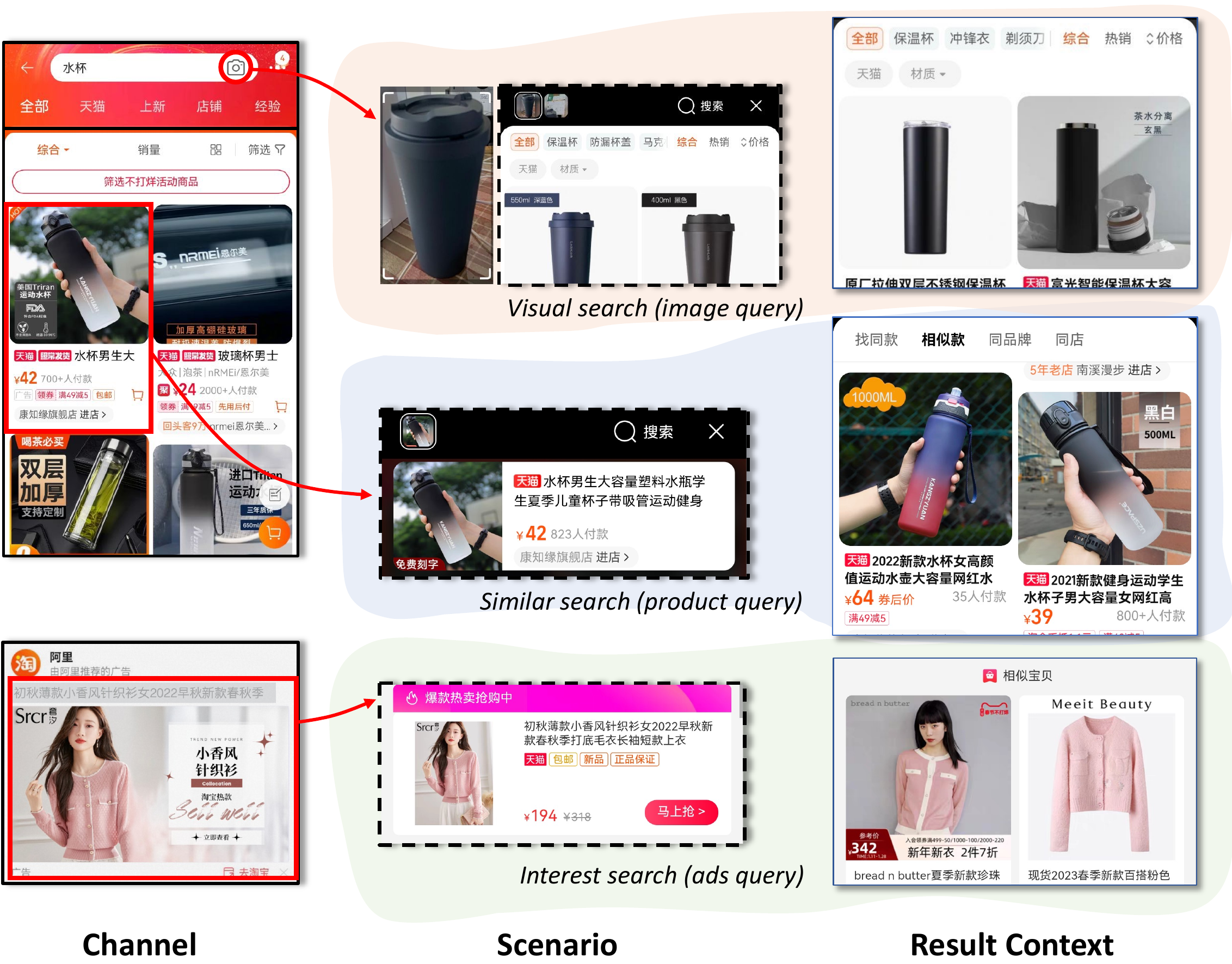}
    \caption{The example of visual search, similar search and interest search.}
    \label{fig:intro1}
\end{figure}

With the rapid development of the Internet, the online business of a company is becoming more diverse and complex. Many different yet potentially relevant scenarios are developed to support various means of information seeking and cover the user's multiple intents.  Figure~\ref{fig:intro1} illustrates three representative business scenarios in Taobao Application: 1) Visual Search~(VS): in addition to keyword search, a user can take a photo containing the desired product for retrieval; 2) Similar Search~(SS): a user plans to retrieve more products that are highly similar to the target product. This time, the user's intention would be relatively more complex, such as price comparison and searching for products in the same style; 3) Interest Search~(IS): This scenario is different from the above scenarios. The advertisement~(ad) as an interest trigger is mainly located in the external media. When a user clicks the ads on an external media, he/she is relocated to the mobile Taobao landing page. Unlike Similar Search, the query about Interest Search combines the ads with user-interested products  of more diversity at the recall stage.

Traditional methods~\cite{cheng2016wide, guo2017deepfm, koren2009matrix, rendle2010factorization, schafer2007collaborative, sarwar2001item, zhou2019deep, zhang2016deep, zhou2018deep, shan2016deep, tian2022multi} serve each scenario with their own data separately, each with a particularly tailored model. However, there are two defects that become more and more worrisome with this strategy:~(1) some scenarios have a small amount of training data, leading to inferior performance;~(2) these scenario-dependent solutions need to be optimized parallelly. These things could become troublesome when upgrading and maintaining in the future. Hence, Multi-Scenario Learning~(MSL) emerges as the times require.

Recently, many methods are proposed to improve the performance of MSL. These efforts focus on how to model the commonalities and distinctions across scenarios at the same time. In general, we can summarize them into three paradigms: Auxiliary Paradigm, Expert Paradigm, and Multi-Tower Paradigm, as demonstrated in Figure~\ref{fourParadigm}~(a,b,c). The shared bottom layer of these models is merely simple, either an MLP layer for dimension reduction or an embedding lookup operation. That is, above the shared bottom layer for feature transformation, different network structures on top of the former are searched. While some of them calls for domain specific expertise\cite{shen2021sar, schoenauer2019multi, sheng2021one}, the others choose to learn the structure automatically\cite{ma2018modeling, tang2020progressive, zou2022automatic}. Though fruitful performance gain is obtained by these solutions, they all take it for granted that the representations generated by the bottom layer are fair for different scenarios.
In a nutshell, it is interesting but unexplored to enable adaptive feature learning for MSL. A Chinese old saying is: a nine-storeyed terrace must be constructed from its very base. Since it is the fatal bottleneck of the whole architecture and hence would eliminate the expensive structure search for the upper layer.

To this end, in this paper, we propose a \textbf{m}ulti-scen\textbf{a}rio \textbf{r}anking framework with adapt\textbf{i}ve fe\textbf{a}ture learning (named \baby). In the bottom layer of \baby, we design three components to enable discriminative feature learning in a scenario-aware manner: \textit{feature scaling}, \textit{feature refinement}, and \textit{feature correlation modeling}. Specifically, we firstly group the  attribute features into different fields like user field, product field, context field and so on. After that, the feature scaling module is utilized to identify the importance of each feature by squeezing or magnifying the feature values. Then, in the feature refinement module, an automatic refiner selection network is utilized for each feature field to perform further high-level semantic encoding. The purpose is to pick the most effective refiner to derive more discriminative semantics at the instance level. Moreover, we further capture the semantic correlation patterns across feature fields as complementary signals. These resultant representations are then concatenated and fed into a simple mixture-of-experts~(MoE) structure. It is worthwhile to mention that the above steps focus on extracting scenario-specific features. Hence, to exploit the shared knowledge, we set up an additional scenario-shared tower for final prediction. The paradigm of our proposed \baby is demonstrated in Figure~\ref{fourParadigm}~(d). 

We perform extensive experiments over two large-scale real-world datasets for product search and recommendation respectively. The results well demonstrate the superiority of the proposed solution against a series of SOTA alternatives. To summarize, the contributions of this paper are as follows:

\begin{figure*}[!]
  \centering
  \begin{subfigure}{0.24\linewidth}
		\centering
		\includegraphics[width=0.9\linewidth]{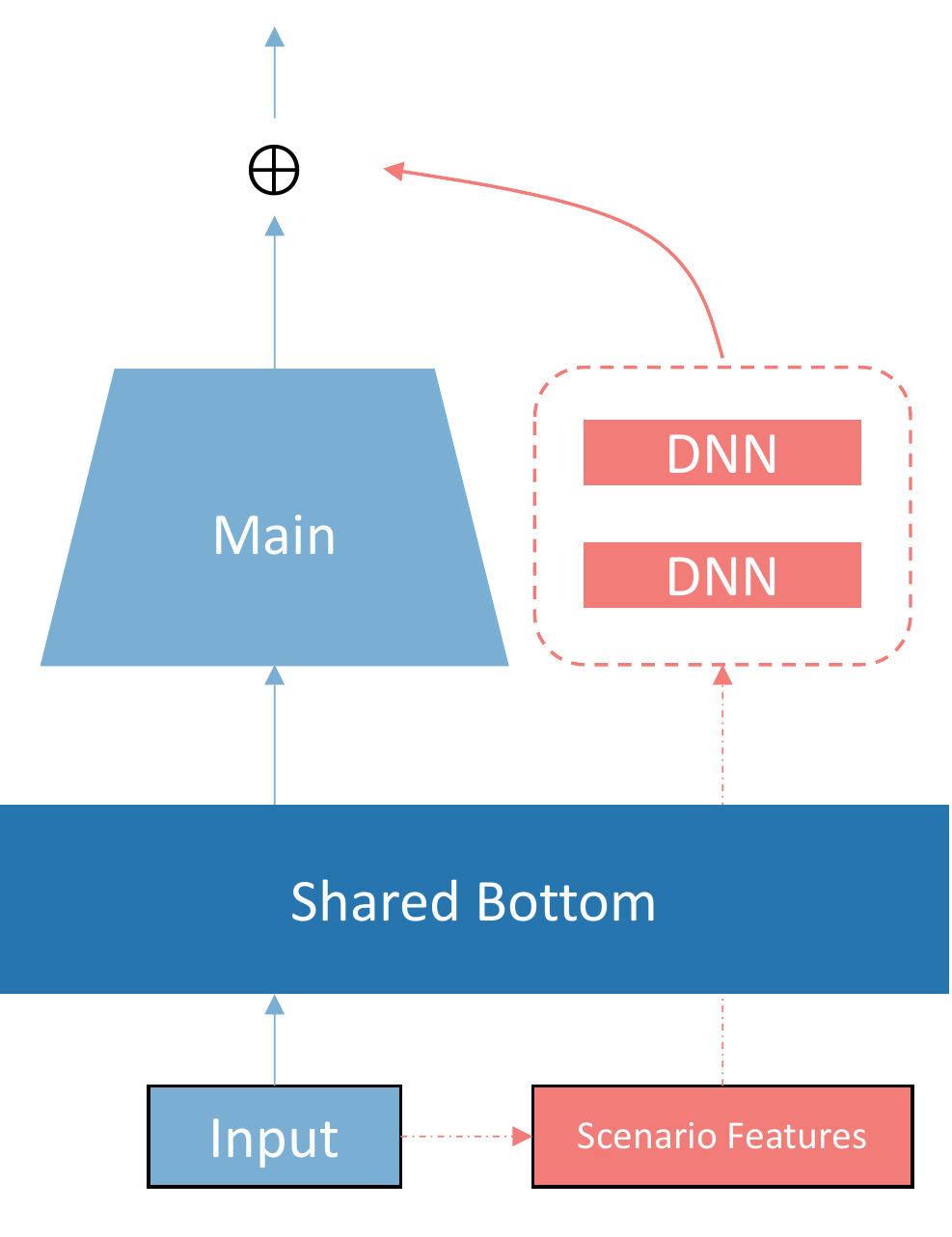}
		\caption{Auxiliary Paradigm}
		\label{paradigm1}
  \end{subfigure}
  \begin{subfigure}{0.24\linewidth}
		\centering
		\includegraphics[width=0.9\linewidth]{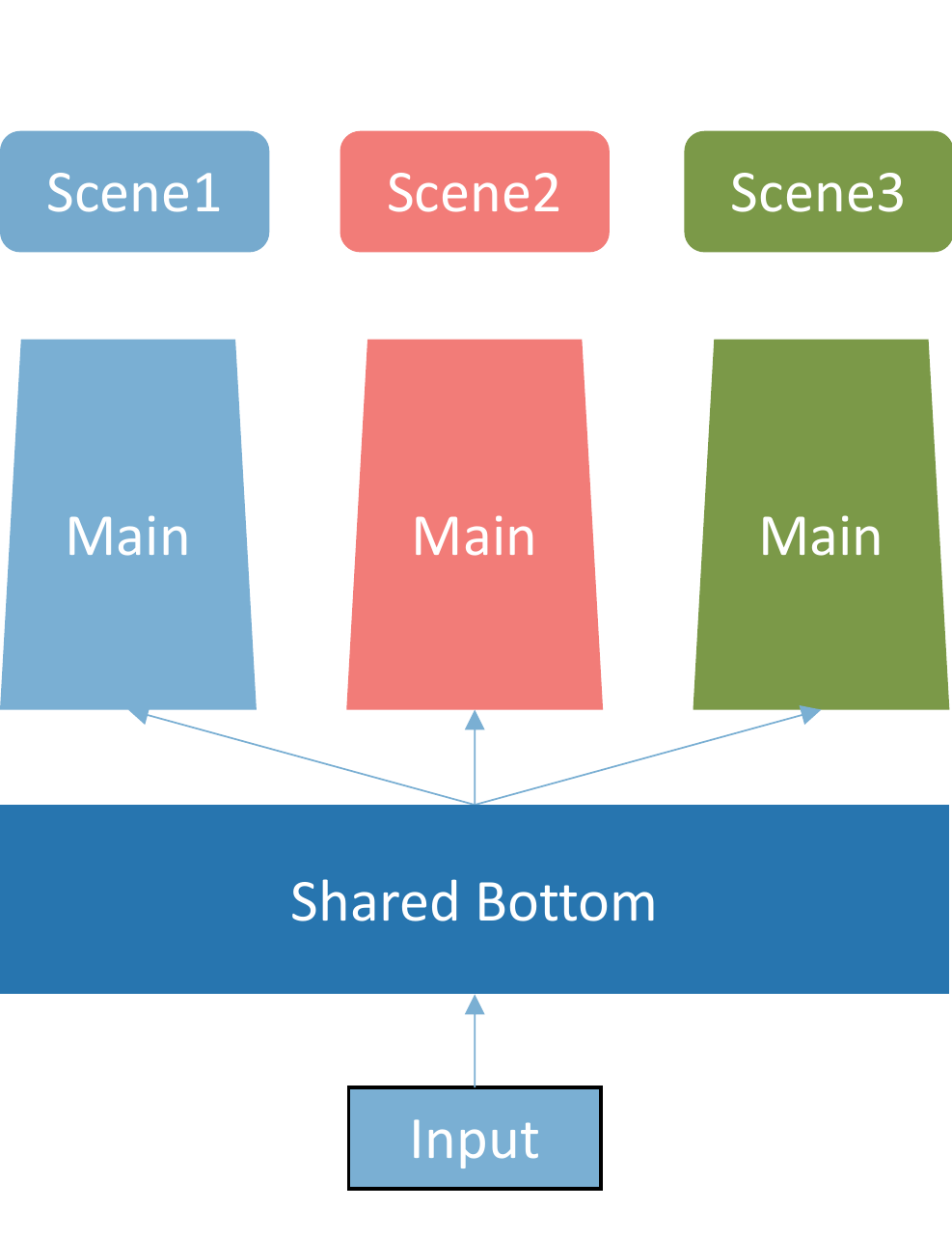}
		\caption{Multi-tower Paradigm}
		\label{paradigm2}
  \end{subfigure}
  \begin{subfigure}{0.24\linewidth}
		\centering
		\includegraphics[width=0.9\linewidth]{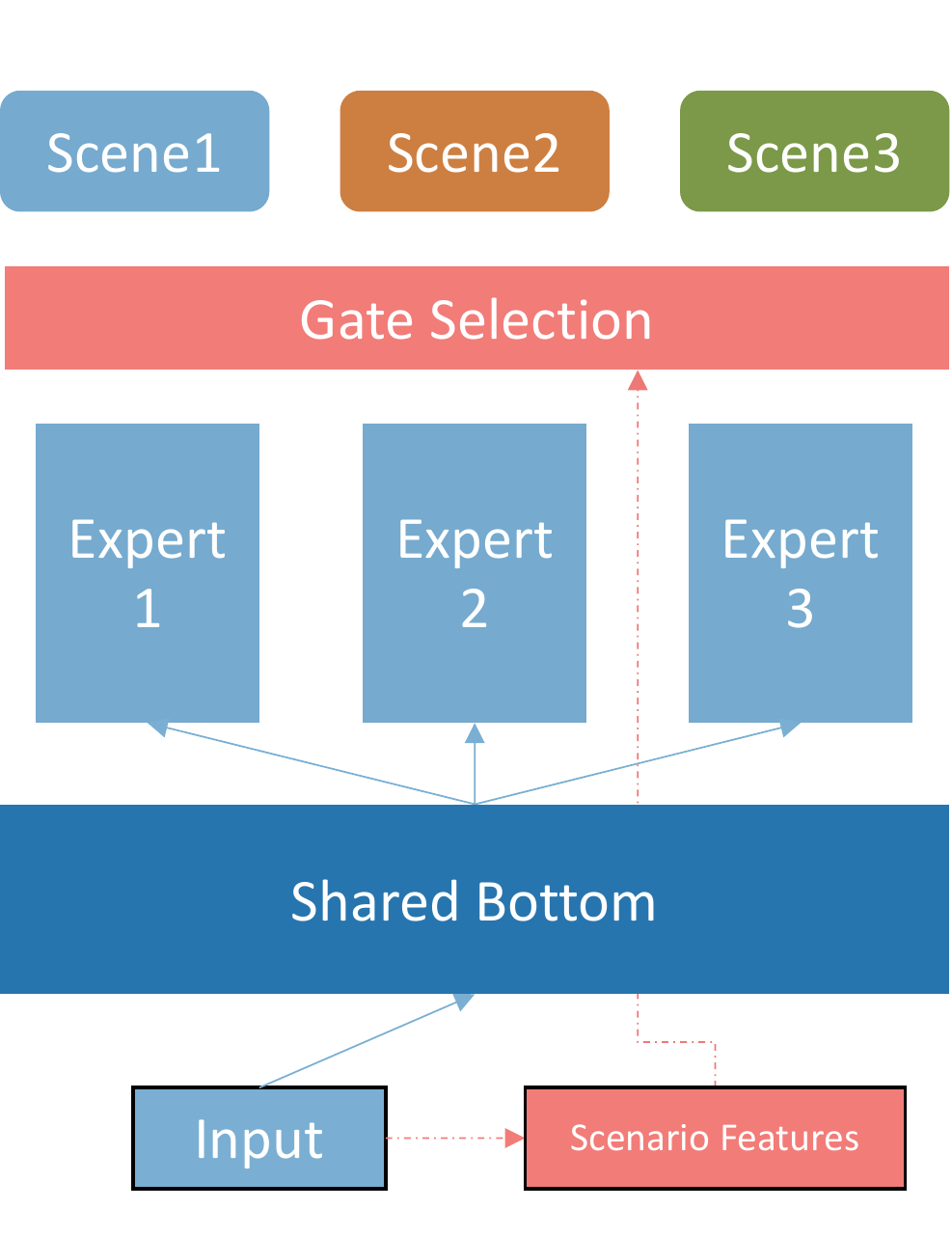}
		\caption{Expert Paradigm}
		\label{paradigm3}
  \end{subfigure}
  \begin{subfigure}{0.24\linewidth}
		\centering
		\includegraphics[width=0.9\linewidth]{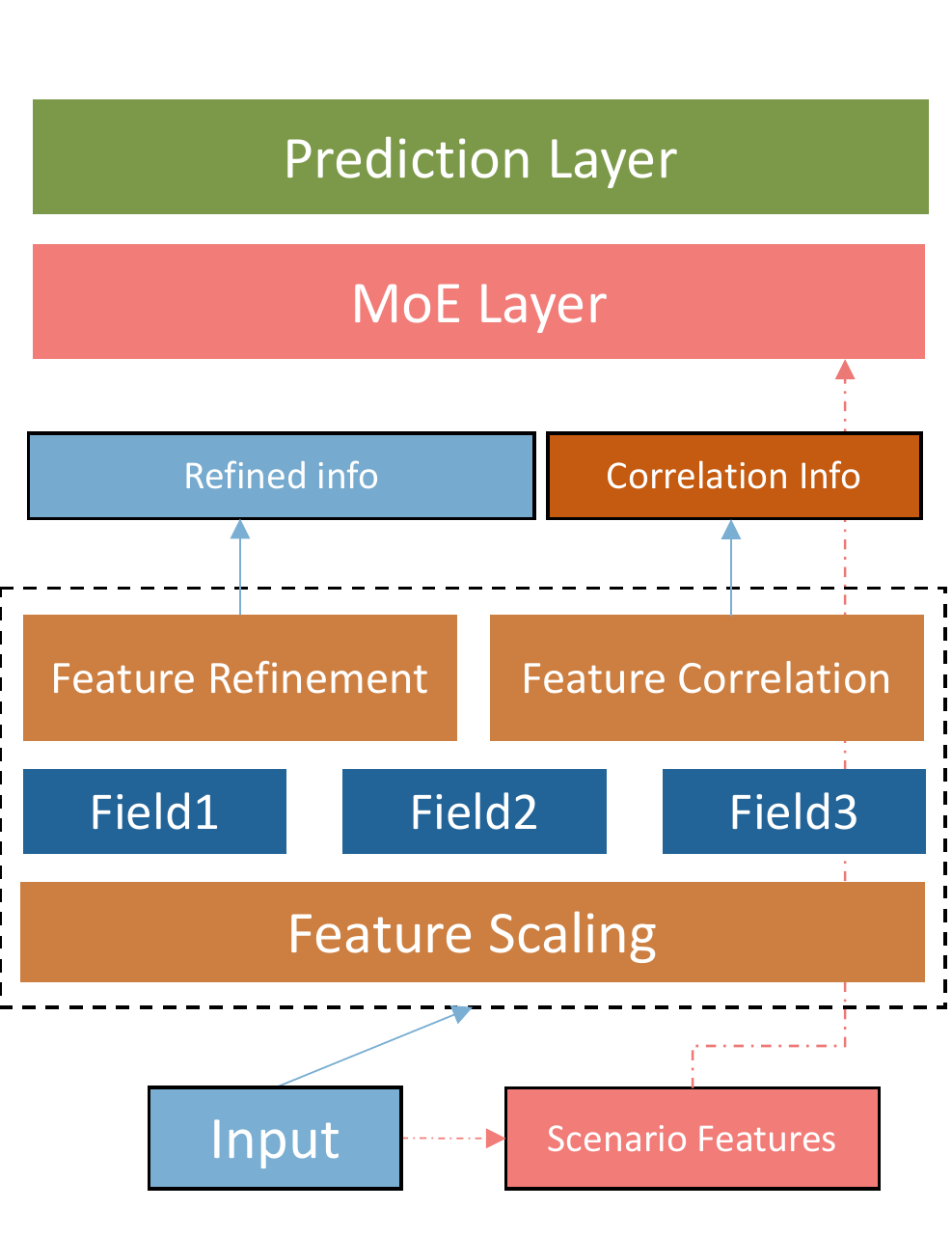}
		\caption{Ours}
		\label{paradigm4}
  \end{subfigure}
  \caption{The current network structures and ours.}
  \label{fourParadigm}
\end{figure*}

\begin{itemize}
    \item We highlight the fact that current MSL researches mainly neglect the significance of performing scenario adaptive feature learning. We argue that learning to extract discriminative features in a scenario-aware manner is simpler yet more effective than expensive network structure optimization.

    \item We propose a multi-scenario ranking framework that is devised with an adaptive feature learning strategy for better retrieval and recommendation. In particular, we introduce three components to derive more discriminative features for the forthcoming multi-scenario learning. 
    
    \item We conduct extensive experiments on two large-scale datasets collected from real-world product search and recommendation application respectively. The experimental results demonstrate that a significant performance gain is obtained by \baby over the state-of-the-art technique alternatives. Further analysis and experiments also validated the cost-effective nature of the \baby.
\end{itemize}
\section{Related Work}

\eat{
Considering both multi-task and multi-scenario learning in recommender and retrieval systems are two major areas related to our work, we therefore briefly summarize the relevant existing methods about these two areas in three paragraphs corresponding to three paradigms. Since some solutions adopt the fusion method of multiple paradigms, we classify them according to the representative modules of these models.
}

\subsection{Auxiliary Paradigm}
Generally, scenario indicators have low dimensions. With the increase in network depth, scenario indicators have a limited impact on the final prediction results. 
Compared with the simplest solutions that use scenario-related features as input, auxiliary solutions build an auxiliary sub-network. In this way, affect the results of the main network by adding or multiplying operation at the output layer directly. This approach is first utilized in single-scenario learning to mitigate the selection bias in training logs, in which many works\cite{AGeneralFrameworkforCounterfactualLearning-to-Rank, joachims2017unbiased, wang2016learning} have been proposed to solve the position bias. These methods usually need to learn a separate model to estimate the influence of bias. Affected by the above methods, DMT\cite{gu2020deep} proposed a bias deep neural network to model neighbor bias and position bias, which aims to improve the performance of multi-objective ranking. And Shen et al~\cite{shen2021sar} found a seesaw phenomenon in multi-task learning, thus, applied the bias network to calculate the deviation of traffic intervention on training according to the information on scenarios. In addition to using auxiliary networks, there are also ways to add the auxiliary loss to enhance specific learning between different scenarios. For instance,  MulANN\cite{schoenauer2019multi} adds a scenario discriminator module to distinguish which scenario the samples are from. 
It can be concluded that the auxiliary paradigm can make the scenario characteristics directly affect the final results, but has limited impact on the intermediate layer of the main network.

\subsection{Multi-Tower Paradigm}
Although the auxiliary paradigm has achieved certain results to some extent, the amount of data in each scenario is seriously unbalanced. In case the whole model will be dominated by data-rich scenarios, multi-tower solutions are proposed by scientists.
Each scenario corresponds to a tower and has its own independent parameter subspace, which further improves the individualization.
The primary model applies shared bottom and a specific Deep Neural Network(DNN) for each scenario. In 2016, Misra et al~\cite{misra2016cross} proposed a Cross-stitch unit to join two independent task networks in an end-to-end learning strategy. Furthermore, DADNN~\cite{he2020dadnn} utilized a routing layer splitting the samples by scenarios into respective domain layers, thus allowing for discriminative representations to be tailored for each individual scenario. More specifically, each scenario has a domain layer that only uses its own data to adjust parameters. Another classical approach STAR~\cite{sheng2021one} added an extra shared tower and proposed a novel parameter fusion method. It multiplies the shared parameters of the extra tower and the customized parameter matrix of the specific scenario to obtain the final network processing parameters. It is worth mentioning that STAR also refers to the first paradigm in scenario information modeling.
It can be concluded that the sharing tower can be added to learn the common information between scenarios and alleviate the problem of poor learning of small scenarios.

\subsection{Expert Paradigm}
In order to enhance the decoupling ability of the model, Google~\cite{shazeer2017outrageously} proposed Mixture-of-Experts~(MoE), which can significantly increase model capacity and capability. This structure is widely used in the MSL area. After that, MMoE~\cite{ma2018modeling} characterizes the task correlation and learns the function of specific tasks based on shared representation. Li et al~\cite{li2020improving} focus on recommendations in multi-country scenarios. The PLE~\cite{tang2020progressive} model shares experts in the share layer and refines tasks uniquely, namely, Customized Gate Control~(CGC) structure. In this way, it effectively alleviates the noise caused by other scenarios and improves the effectiveness of feature extraction. Besides, Zhang et al~\cite{zhang2022leaving} utilize expert networks to solve multi-scenario and multi-task problems on the advertiser's side and use the meta network to express the scenario information explicitly. Zou et al~\cite{zou2022automatic} propose a novel expert network structure with automatic selection of fine granularity. By calculating the KL divergence of the gated column vector, one-hot vector, and uniform vector, gates can select the most suitable sharing and exclusive experts. 
To summarize, the expert paradigm borrowed the idea of bagging, more specifically, training multiple experts to make decisions. This decision is better than one expert in terms of generalization, expression, and learning ability. At the same time, the setting of gates increases flexibility, which takes into account the different learning modes of different scenarios.

\section{Method}\label{sec:algo}
\begin{figure*}[htbp]
\centerline{
\includegraphics[width=0.95\textwidth]{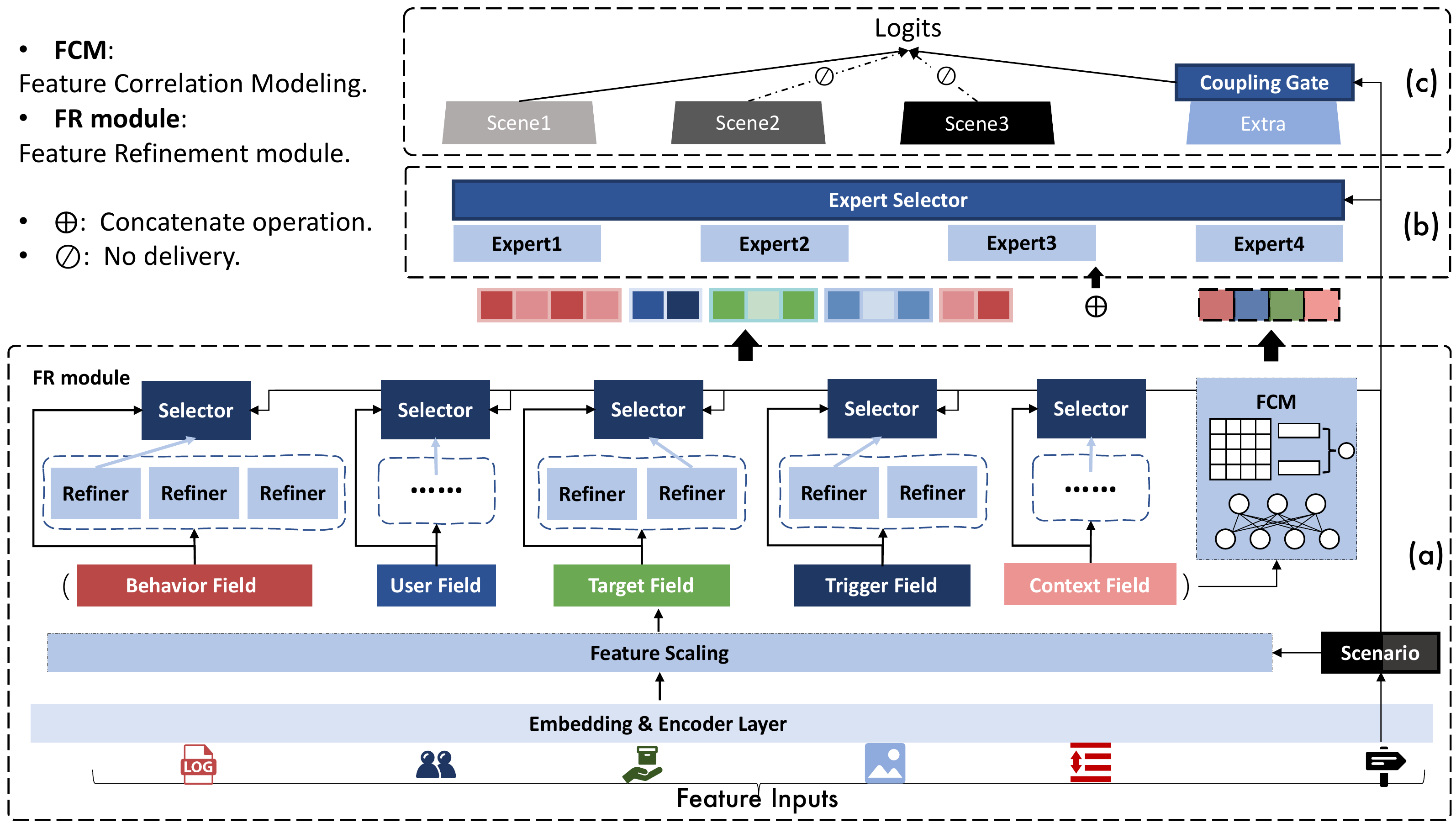}}
\centering
\caption{The network architecture of our proposed \baby.}
\label{fig:model}
\end{figure*}

In this section, we present the proposed \baby model in detail. Figure~\ref{fig:model} illustrates the network architecture of our model.

\subsection{Problem Formulation}\label{ssec:formulation}

Let $\mathcal{V},\mathcal{U},\mathcal{S}$ be the set of all products~(including ads), all users, and all scenarios available respectively, and $\{b_u\}_{M}^N$ be the behavior sequence set where $M=|\mathcal{V}|, N=|\mathcal{U}|$ and $S=|\mathcal{S}|$. Given each user $u$, the corresponding behavior sequence is organized by following the chronological order: $b_u=[x_1,x_2,\cdots,x_m]$, where $x_j\in\mathcal{V}$ for $1\le j\le m$, and $m$ is the predefined maximum capacity. Product field contains $P$ attributes: $\mathcal{A}_x = [a_x^1,a_x^2,\cdots,a_x^P]$, while user field contains $L$ attributes: $\mathcal{A}_u = [a_u^1,a_u^2,\cdots,a_u^L]$. As to the attributes having continuous numbers, we transform the value into categorical features. Similarly, let $\mathcal{C}_{us}=\{c_u^1, c_u^2, ..., c_u^{N_c}\}$ denote $N_c$ context attributes associated with each user $u$ under scenario $s$. These context features usually describe the page and physical environments of the user side. The total number of available attribute features for users, products and contexts are denoted as $N_a^u$, $N_a^x$, and $N_a^c$ respectively. Our task is to precisely identify the target product $x_{i}$ that will be interested by user $u$ with the above knowledge. 

Note that our purpose is to devise a unified ranking framework applicable to both product search and sequential recommendation. For product search, we have an additional input query provided from the user. Since those $S$ scenarios cover diverse services, queries could be in various heterogeneous forms including images, products, and ads, besides the texts. Hence, to keep a concise description in the rest parts, we use the notion of \textit{trigger} to denote the queries under different application scenarios without further clarification. Moreover, without loss of generality, we can assume that each trigger $t$ also includes $O$ attributes: $\mathcal{A}_t=[a_t^1,a_t^2,\cdots,a_t^O]$ and $N_a^t$ denotes the total feature number accordingly. For target scenario $s$ and the corresponding trigger $t$, a model $f()$ is optimized to precisely generate the likelihood for each candidate product as follows:

\begin{align}
\hat{y}_{ui}=f(x_i\mid b_u,s,t,\mathcal{C}_{us},\mathcal{A}_u,\mathcal{A}_t,\{\mathcal{A}_{x_1},\mathcal{A}_{x_2},\cdots,\mathcal{A}_{x_m}\});
\end{align}
where $\hat{y}_{ui}$ is the likelihood that user $u$ will interact with product $x_i$ with respect to trigger $t$\footnote{The trigger will not exist under recommendation scenario. We consider the trigger field when describing the algorithm.}.

\subsection{Encoder Layer}\label{ssec:encoderlayer}
\paratitle{Embedding Layer.} 
In the embedding layer, we firstly utilize seven embedding tables to encode the semantics for users, products, user attributes, product attributes, queries, query attributes, contexts, and scenarios respectively: $\mathbf{U}\in R^{N\times d_u},\mathbf{V}\in R^{M\times d_x},\mathbf{A}_u\in R^{N_a^u\times d_a},\mathbf{A}_x\in R^{N_a^x\times d_a},\mathbf{T}\in R^{{N_a^t}\times d_a},\mathbf{C}\in R^{N_a^c\times d_c},\mathbf{S}\in R^{S\times d_s}$, where $d_u$, $d_x$, $d_a$, $d_c$, and $d_s$ denote the embedding size of the user, product, attribute, trigger, context, and scenario features respectively. Hence, for each product, user, attribute, trigger, context, and scenario, we can perform the table lookup from these matrices to obtain the corresponding embedding representations. 

Then, we concatenate the product embedding and the corresponding attribute embeddings to fully represent the semantics of the product:

\begin{align}
\mathbf{x}=[\mathbf{e}_x||\mathbf{a}_x^1||...||\mathbf{a}_x^P]
\label{equ:X}
\end{align}
where $\mathbf{e}_x$ is the product embedding for product $x$, $\mathbf{a}_x^j$ is the corresponding attribute embedding for $j$-th attribute $a_x^j$, and $||$ denotes the vector concatenation. Consequently, we can form the feature matrix $\mathbf{B}_u=[\mathbf{x}_1,\mathbf{x}_2,\cdots,\mathbf{x}_m]$ for sequence $b_u$. Similarly, we represent the semantics of a user, a trigger, and the corresponding context as $\mathbf{u}=[\mathbf{e}_u||\mathbf{a}_u^1||\cdots||\mathbf{a}_u^L]$, $\mathbf{t}=[\mathbf{\ast}||\mathbf{a}_t^1||\cdots||\mathbf{a}_t^{O}]$ and $\mathbf{c}=[\mathbf{c}_u^1||\cdots||\mathbf{c}_u^{N_c}]$ respectively, where $\mathbf{\ast}$ denotes $\mathbf{x}$ for product based trigger or image embedding for the visual trigger. 
Here, we use linear projection to fix the embedding size of the product-based trigger to be the same as the image embedding, which is marked as $d_t$. The image embedding is pretrained based on the Alimama platform.

\paratitle{Sequence Encoder.} As for the user's consecutive behaviors modeling, we choose to extract the user's preference signals from $b_u$ with a Transformer encoder. The built-in stacked self-attention mechanism could enable us to further exploit semantic correlations across different scenarios. In detail, we apply the Transformer network on top of $\mathbf{B}_u$ as follows:
\begin{align}
 \mathrm{H}_u = [\mathbf{h}_{x_1},\mathbf{h}_{x_2},\cdots,\mathbf{h}_{x_m}] = Trans(\mathbf{B}_u),
\end{align}
where function $Trans$ denotes the Transformer network and the details can be referred to \cite{vaswani2017attention}, $\mathbf{h}_{x_j}$ is the sequence-wise representation for product $x_j$ from the inter-scenario perspective. After that, a trigger-aware attention module is utilized to aggregate the relevant features towards trigger $t$ as follows:

\begin{align}
    \mathbf{h}_{b} &= \sum_{i=1}^{m} \alpha_i\mathbf{h}_{x_i}\\
    \alpha_i &= \frac{\exp\big(sim(\mathbf{t}, \mathbf{h}_{x_i})\big)}{\sum_{j=1}^{m}\exp\big(sim(\mathbf{t}, \mathbf{h}_{x_j})\big)}\\
    sim(x,y) &= FC([x||y]).
\end{align}
where $\mathbf{h}_{b}$ is the trigger-aware representation of the user's historical behaviors, and function $FC()$ denotes the fully-connected layer. In the recommendation scenario, the trigger representation $\mathbf{t}$ is replaced by the target representation $\mathbf{x}_i$ referring to Equation~\ref{equ:X}. Afterwards, we concatenate the representations of different fields together as input for later steps:

\begin{align}
\mathbf{Q}=[\mathbf{h}_b||\mathbf{u}||\mathbf{x_i}||\mathbf{t}||\mathbf{c}],
\label{equ:Q}
\end{align}
where $\mathbf{x_i}$ is the representation of target product $x_i$ field according to Equation~\ref{equ:X}.

\subsection{Feature Scaling}\label{ssec:fs}
Considering the heterogeneity of different scenarios, such as different triggers, it is apparent that different kinds of features have different discrimination in various scenarios. For example, visual features generally play a more significant role in the photo query scenario than that in the product query scenario. What's more, we also consider the interplay between the user's interest and the target product.

Thus, the FS module is designed to squeeze or magnify each feature according to the scenario indicator. Specifically, we calculate a scaling factor for each feature element in $\mathbf{Q}$. Here, a feature element refers to an individual representation. For example, according to Equation~\ref{equ:Q}, $\mathbf{h}_b$ is a feature element, $\mathbf{e}_x$ and $\mathbf{a}_x^1$ are also feature elements. Let $N_Q$ denotes the total number\footnote{$N_Q=L+P+O+N_c+4$} of feature elements included in $\mathbf{Q}$, we calculate a scaling vector $\boldsymbol{\alpha}\in R^{N_Q}$ as follows:
\begin{align}
   \boldsymbol{\alpha} = \lambda*sigmoid(FCN([freeze(\mathbf{Q})||\mathbf{e}_u||\mathbf{e}_{x_i}||\mathbf{e}_s])) 
\end{align}
where $freeze()$ denotes the stop gradient operation which aims to avoid the overfitting phenomenon and gradient conflict problem~\cite{sener2018multi}, $\mathbf{e}_s$ is the scenario embedding. Compared with the standard attention mechanism, the scaling factor $\lambda$ can be larger than one. That is, we can magnify the important signals instead of keeping them remain. Afterwards, the output of FS module $\mathbf{Q}_{S}$ is generated as follows:
\begin{align}
    \mathbf{Q}_{S} &= [\boldsymbol{\alpha}_1\mathbf{Q}_1||...||\boldsymbol{\alpha}_{N_Q}\mathbf{Q}_{N_Q}]\\
    &=[\boldsymbol{\alpha}_1\mathbf{h}_b||\boldsymbol{\alpha}_2\mathbf{e}_u||...||\boldsymbol{\alpha}_{N_Q}\mathbf{c}_u^{N_c}]
\end{align}
where $\boldsymbol{\alpha}_j$ refers to $j$-th scalar value in vector $\boldsymbol{\alpha}$, and $\mathbf{Q}_j$ refers to $j$-th feature element in the order according to Equation~\ref{equ:Q}. We can also denote the scaled representations of different fields as $\hat{\mathbf{h}}_{b},\hat{\mathbf{u}},\hat{\mathbf{x}}_i,\hat{\mathbf{t}},\hat{\mathbf{c}}$ and $\mathbf{Q}_{S}=[\hat{\mathbf{h}}_{b},\hat{\mathbf{u}},\hat{\mathbf{x}}_i, \hat{\mathbf{t}},\hat{\mathbf{c}}]$. Note that scaling vector $\boldsymbol{\alpha}$ is calculated by taking all available features of these five fields into account, which implicitly models the feature correlations. 

\subsection{Feature Refinement}\label{ssec:FR}
To accommodate scenario-specific features at the instance level, we design a feature refinement module, which utilizes an automatic refiner selection network to support high-level semantic encoding. Specifically, we set up for each feature field a set of feature refiners, each of which is a shallow fully-connected layer. As shown in Figure~\ref{fig:model}(a), the selection of refiners is automatically made in a scenario-aware manner. A selector is utilized to calculate weights $\boldsymbol{\beta}$ for each field that takes scenario embedding $\mathbf{e}_s$ and the field representation as input. Taking the user behavior field $\hat{\mathbf{h}}_b$ as an example, the high-level feature vector is calculated as follows:
\begin{align}
    \boldsymbol{\beta} &= GS(sigmoid(FCN([\hat{\mathbf{h}}_b || \mathbf{e}_s]))),~~\boldsymbol{\beta}\in R^{N_{b}} \\ 
    \Tilde{\mathbf{h}}_{b} &=  [\boldsymbol{\beta}_1 FC_1(\hat{\mathbf{h}}_b)||\cdots||\boldsymbol{\beta}_{N_b}FC_{N_b}(\hat{\mathbf{h}}_b)]
\end{align}
where $GS$ and $sigmoid$ denote Gumbel Softmax layer and sigmoid activation respectively, $N_b$ denotes the number of refiners deployed for this feature field, $FC_j()$ denotes the $j$-th refiner for the field, and $\Tilde{\mathbf{h}}_b$ is the resultant high-level feature vector. Here, $GS$ layer is used to approximate the discretization nature of the selection process. We use $ReLU$ as the activation function for each refiner. The same procedure is performed for other feature fields, the resultant high-level features can be represented as follows:
\begin{align}
    \mathbf{Q}_{R} = [\Tilde{\mathbf{h}}_b||\Tilde{\mathbf{u}}||\Tilde{\mathbf{x}}_i||\Tilde{\mathbf{t}}||\Tilde{\mathbf{c}}].
\end{align}

\subsection{Feature Correlation Modeling}\label{ssec:FCM}
After feature scaling, we further choose to explicitly model the semantic correlations across different feature fields. At first, the representation of a field is projected into the same dimension size $d_r$ via an independent fully-connected layer. The resultant field representations are denoted as $\bar{\mathbf{h}}_b, \bar{\mathbf{u}}, \bar{\mathbf{x}}_i, \bar{\mathbf{t}}, \bar{\mathbf{c}}$. Then, we calculate the dot product for each field pair, and concatenate the scores as follows:
\begin{align}
    \mathbf{Q}_C = [\bar{\mathbf{h}}_b\cdot\bar{\mathbf{u}}||\bar{\mathbf{h}}_b\cdot\bar{\mathbf{x}}_i||\cdots||\bar{\mathbf{t}}\cdot\bar{\mathbf{c}}]
\end{align}
where symbol $\cdot$ denotes the dot product between two vectors. Ultimately, the final output of the adaptive feature learning by our \baby is formed by concatenating both $\mathbf{Q}_{R}$ and $\mathbf{Q}_{C}$:
\begin{align}
    \mathbf{Q}_{f} = [\mathbf{Q}_{R}||\mathbf{Q}_{C}].
\end{align}

\subsection{Network Layer}\label{ssec:nld}
After the adaptive feature learning described above, we utilize a standard Mixture-of-Experts~(MoE)~\cite{shazeer2017outrageously} as the main structure in the network layer (as shown in Figure~\ref{fig:model}~(b)). Briefly, we set up a shared set of experts for all $S$ scenarios, where an expert is an independent fully-connected network. A gating mechanism is utilized to aggregate the output of these experts as follows:
\begin{align}
    \mathbf{h}_{N} &= \sum_{j=1}^{N_{e}}\mathbf{g}f_j(\mathbf{Q}_{f})\\
    \mathbf{g} &= softmax(\mathbf{W}_g\mathbf{e}_s)
\end{align}
where $N_{e}$ denotes the number of experts in network layer, $W_g$ is a learnable parameter of the gating mechanism, vector $\mathbf{g}$ contains the importance weights of the experts, and $f_j$ is the $j$-th expert. In detail, we use $ReLU$ as the activation function.

\subsection{Prediction and Model Optimization}\label{ssec:predictor}
\paratitle{Prediction.}
In the prediction layer, we draw lessons from the advantages of multi-tower structure and introduce the scenario-specific DNN tower $FCN_{sp}^{s}(·)$ in the prediction phase. In addition, an extra tower $FCN_{sh}(·)$ is utilized to harness scenario-shared information (as shown in Figure~\ref{fig:model}~(c)). Thus, the final representation $\mathbf{h}_{f}$ used for prediction is obtained from these two perspectives:
\begin{align}
    \mathbf{h}_{f} &= \mathbf{h}_{sp}^s + \alpha_{s}\mathbf{h}_{sh}\\
    \mathbf{h}_{sp}^s &= FCN_{sp}^{s}(\mathbf{h}_{N})\\
    \mathbf{h}_{sh} &= FCN_{sh}(\mathbf{h}_{N})
\end{align}
where $\alpha_s$ is a coupling coefficient to control the impact of the shared information. Here, we expect that $\alpha_s$ should be small when the target scenario has little connection to other scenarios. Therefore, we calculate $\alpha_s$ by measuring the relevance across scenarios as follows:
\begin{align}
    \alpha_{s} = \frac{1}{N_s-1}\sum_{j=1,s_j\neq 
 s}^{N_s}\mathbf{e}_{s}\cdot\mathbf{e}_{s_j}
\end{align}

Finally, the likelihood that user $u$ will interact with product $x_i$ is calculated as follows:
\begin{align}
\hat{y}_{ui} = FCN(\mathbf{h}_{f}),
\label{eq:score}
\end{align}
where $\hat{y}_{ui}$ denotes the prediction score, $FCN()$ denotes the fully-connected network with sigmoid activation.

\paratitle{Model Optimization.}
For the sake of ensuring the high mobility and universality of the model, all samples use the cross entropy loss, although there are multiple scenarios in the dataset. Thus, the final loss is formulated as follows:
\begin{align}
    \label{eq:finalloss}
    \mathcal{L}_{final} &= \mathcal{L}_1 + \gamma\mathcal{L}_2, \\
    \mathcal{L}_1 &= -\sum_{u,i}[y_{ui}ln(\hat{y}_{ui}) + (1-y_{ui})ln(1-\hat{y}_{ui})],
\end{align}
where $y_{ui}, y_{ui}\in \{0, 1\}$ denotes the ground truth, $\mathcal{L}_2$ is the $L_2$ norm of all model parameters, $\gamma$ is a hyperparameter to control the former.

\begin{table}[t]
\small
\setlength{\tabcolsep}{1pt}
\centering
    \caption{Statistics of the ALi-CCP datasets. \#product, \#user \#impression and \#click represent the number of products/ads, users, user views, clicks respectively.}
    \label{tab:datastats}
    \begin{tabular}{c|cccc}
    \hline
    \textbf{Datasets}&\multicolumn{4}{c}{Ali-CCP}\\
    \hline
    \textbf{Scenarios}&S1&S2&S3&All\\
    \hline
    \textbf{\#User}&91,488&2,612&154,024&244,397\\
    \textbf{\#Product}&535,711&198,651&537,937&538,376\\
    \textbf{\#Impression}&32,236,951&639,897&52,439,671&85,316,519\\
    \textbf{\#Click}&1,291,063&28,022&1,998,618&3,317,703\\
    \bottomrule
  \end{tabular}
\end{table}


\section{Experiments}
In this section, we conduct extensive experiments over two large-scale real-world datasets to validate the efficacy of \baby in both item search and item recommendation scenarios. 

\subsection{Experimental Settings}
\paratitle{Datasets.} To validate the efficacy of our proposed \baby, two real-world large-scale datasets covering diverse scenarios are used for performance evaluation. The first one is the Alimama retrieval dataset collected from the Alibaba search advertising platform. As aforementioned in Section~\ref{sec:intro}, this dataset covers daily search logs of the three scenarios in the period of 2022/08/25 to 2022/09/23: 1) Visual Search~(VS) based on the taken photo; 2) Similar Search~(SS) for relevant product search against the selected target product; and 3) Interest Search~(IS) for ads search from external traffic. We take the instances of the last day, the penultimate day, and earlier ones for testing, validation, and training respectively. 

The second dataset is Ali-CCP\footnote{https://tianchi.aliyun.com/dataset/408}, which is widely used in the relevant literature~\cite{li2020improving, ma2018entire} for product recommendation, which was collected Taobao’s recommender system under three scenarios. However, the whole dataset has been anonymized, we cannot describe the specific business scenario. Instead, we denote these scenarios as S1, S2, and S3. The training set, validation set, and test set based on the official split are taken for experiments~\cite{ma2018entire}.

Table~\ref{tab:datastats} reports the statistics of Ali-CCP dataset in detail. Due to the sensitive business information involved in the data, we are allowed to report the detailed numbers. \#product, \#user \#impression and \#click is at hundred millions, ten millions, billions and hundred millions level respectively. 
To keep the true characteristics of the real-world scenarios, we do not include further preprocessing steps. In terms of count numbers, we can see that the Alimama dataset is at least two orders of magnitude larger than Ali-CCP, though both of them contain user behaviors of large-scale. Furthermore, there are also obvious distribution discrepancies between scenarios. For example, the number of impressions in the SS scenario of the Alimama dataset is much smaller than the other two scenarios. The same imbalance also occurs in the Ali-CCP dataset. To summarize, these datasets have different data characteristics and distributions covering a broad range of real-world situations, which can effectively verify the effectiveness and universality of our model.

\begin{table*}[!]
\centering
\caption{Performance comparison of different methods across the two datasets. Each dataset consists of three scenarios respectively. The best and second-best results are highlighted in boldface and underlined respectively. $*$ indicates that the performance difference against the best result is statistically significant at $0.05$ level.}
    \label{tab:results}
    \begin{tabular}{c|c|ccccccc|c}
    \toprule
    \multirow{2}{*}{Dataset}&\multirow{2}{*}{Scenarios}&\multicolumn{8}{c}{Models}\cr
    \cmidrule(lr){3-10}
    & &Hard Sharing&Shared Bottom &MulANN&MMoE&PLE&STAR&$AEMS^2$&\baby\cr
    \cmidrule{1-10}
    \multirow{4}{*}{Alimama}& VS & 0.7415* & 0.7318* & 0.7423* & \underline{0.7441}*  & 0.7421* & 0.7323* & 0.7289*  & \textbf{0.7473} \cr
    & SS & 0.6777* & \underline{0.6842}* & 0.6792* & 0.6779*  & 0.6840* & 0.6724* & 0.6703* & \textbf{0.6927} \cr
    & IS & 0.7131* & \underline{0.7159}* & 0.7129* & 0.7139*  &0.7126* & 0.6925* & 0.6994* & \textbf{0.7178} \cr
    \cmidrule{2-10}
    & Total Impr. & +0.0255 & +0.0259 & +0.0234 & +0.0219 & +0.0191 & +0.0606 & +0.0592 & -- \cr
    \midrule
    \multirow{4}{*}{Ali-CCP}& S1 &  0.5747* & 0.5528* & 0.5625* & 0.5740*  & 0.5744* & 0.5524* & \underline{0.5773}* & \textbf{0.5869} \cr
    & S2   & 0.5912* &0.5606* & 0.5779* & 0.5789*   & 0.5939* & 0.5826* & \underline{0.5957}* & \textbf{0.6114} \cr
    & S3   & 0.5888* &0.5710* & 0.5754* & 0.5768*  & 0.5927* & \underline{0.5975}* & 0.5916* & \textbf{0.6077} \cr
    \cmidrule{2-10}
    & Total Impr. & +0.0513 & +0.1216 & +0.0902 & +0.0763 & +0.0450 & +0.0735 & +0.0414 & -- \cr
    \bottomrule
    \end{tabular}
\end{table*}

\paratitle{Baselines.}
We compare the proposed \baby\footnote{The code implementation is available at https://github.com/WHUIR/Maria.} against the following state-of-the-art methods:
\begin{itemize}[leftmargin=-0.1em]
    \item \textbf{Hard Sharing} is a multi-scenario model that shares the parameters of the bottom layer. On top of the shared bottom layer, a single DNN is used for prediction across scenarios. 
    \item \textbf{Shared Bottom (Multi-DNN)} replaces the single DNN with multiple DNNs. That is, an individual DNN is utilized for each scenario. And we add an auxiliary tower to enhance the ability to characterize the scenario indicator.
    \item \textbf{MulANN}~\cite{schoenauer2019multi} introduces a domain discriminator module, which aims to distinguish which scenario the instance is from. Adversarial learning is utilized to avoid overfitting against the scenario-specific features, leading to better scenario-shared knowledge transfer.
    \item \textbf{MMoE}~\cite{ma2018modeling} implicitly models task relationships for multi-task learning, where different tasks may have different label spaces. Here we adapt MMoE for multi-scenario learning. The number of experts is equal to the number of experts of \baby. The sum of weighted outputs from the experts are fed into the individual tower for each scenario respectively.
    \item \textbf{PLE}~\cite{tang2020progressive} is a state-of-the-art multi-scenario/multi-task model that organizes the experts into scenario-specific groups and scenario-shared groups for the purpose of avoiding negative transfer or seesaw phenomenon.
    \item \textbf{STAR}~\cite{sheng2021one} proposes a star topology to accommodate with the scenario-specific characteristics. Specifically, a shared network works as the center node for knowledge sharing and each scenario network connects only with the center node.
    \item \textbf{AEMS}$^2$~\cite{zou2022automatic} proposes a novel MMoE-based model with automatic search towards the optimal network structure. In contrast to PLE and STAR, an expert can be either scenario-shared or scenario-specific dynamically in an instance-aware manner. 
\end{itemize}

\paratitle{Hyperparameter Settings.}
For a fair comparison, all methods are implemented in the Tensorflow framework, and Adam optimizer is utilized with default parameter setting. Moreover, the number of experts $N_e$ is set to $4$ in an expert layer. The learning rate, mini-batch size, and decay rate are set to $0.05$, $512$, and $1e^{-2}$ respectively. Moreover, we set the number of hidden units to $256$ for each expert that is instantiated with a two-layer fully-connected network. A single expert layer is used for all MoE-based models, except for PLE where two expert layers are stacked. That is, we aim to keep their unique network structures but keep the model sizes comparable. As for the prediction layer, a DNN with $128\times64\times32\times1$ structure is adopted for baselines and the towers in \baby. The weight of domain loss in MulANN is $0.1$ on the Ali-CCP and $0.01$ on the Alimama dataset respectively. 

As to \baby, we find our model performs relatively stable when $\lambda$ is set to $2$, and the temperature is set to $0.01$ for the Gumbel Softmax layer. Also, for the Ali-CCP dataset, $\gamma$ is set to $1e^{-2}$ and the number of refiners is set to $2$ for the user field, while a single refiner is used for the rest fields. For the Alimama dataset, these values are set to $1e^{-6}$ and $2$ respectively. Moreover, the dimension size of the refined field representation is reduced against the counterpart in Equation~\ref{equ:Q}. Because we observe little performance fluctuation in a wide range of compression ratio, say $50\%$-$85\%$.

\paratitle{Evaluation Metric.}
We adopt the area under the ROC curve (AUC) for performance evaluation. For each method, we repeat the experiment five times and report the averaged results. The statistical significance test is conducted by the student $t$-$test$.

\subsection{Performance Evaluation}
The overall performance of all methods is reported in Table~\ref{tab:results}. Here, we make the following observations.

As for traditional Hard Sharing, it is very difficult to achieve satisfied performance. Compared with other models with expert network or multi-tower structure, the hard sharing solution is not suitable for complex multi-scenario learning. Shared Bottom~(Multi-DNN) adds a specific DNN for each scenario in order to harness the scenario-specific knowledge. However, this simple strategy performs suboptimal in two scenarios on the Alimama dataset but performs poorly on the Ali-CCP dataset. This phenomenon is caused by the different characteristics of the two datasets. The scenarios of the Alimama are quite different, as aforementioned. On the contrary, the scenarios in the Ali-CCP dataset are relatively more similar to each other. Note that the advantage of Shared Bottom over Hard Sharing is reversed on Ali-CCP. This suggests that these two trivial methods are inferior to handling complex MSL.

By reinforcing the extraction of the scenario-shared knowledge, MulANN seems to achieve better performance than the above two baselines. But, it is obvious to observe the seesaw phenomenon across scenarios. Moreover, it is surprising that MMoE achieves the second-best performance in the VS scenario on the Alimama dataset. However, the data imbalance problem can not be well addressed by MMoE, leading to inferior performance in the SS scenario instead. As a variant of MMoE, PLE separates the experts into two groups, which alleviates the problems of MMoE somehow. The performance of PLE in various scenarios is more stable. We can see that in the Ali-CCP dataset, PLE shows a significant improvement over MMoE. Furthermore, AEMS$^2$ performs relatively well on the Ali-CCP dataset but worse on the Alimama. That is, AEMS$^2$ is possibly more desired for scenarios that are similar to each other. Also, the STAR model only achieves suboptimal performance in S3 on Ali-CCP. Overall, we can see that there is no dominating network structure that can handle complex MSL tasks in real-world scenes.

Our proposed \baby has significant yet consistent improvement across the three scenarios and two datasets. Specifically, we also summarize the total improvement over each dataset by summing the relative performance gain over the three scenarios. The relative improvement is up to $6.06\%$ and $12.16\%$ for Alimama and Ali-CPP datasets respectively. Recall that our network layer only counts on a standard MoE structure. These results suggest that the idea of performing adaptive feature learning is critical for better MSL.

\subsection{Model Analysis}
In this section, we perform a deep analysis of our \baby. At first, the model complexity analysis is conducted to illustrate that the \baby gains positive benefits with comparable computation complexity and the model size against these baseline methods. Then, a series of ablation studies are performed to validate each design choice. At last, we further dive into the working mechanism of automatic refiner selection with some visualizations.

\paratitle{Complexity Analysis.} Since the adaptive feature learning part is unique for \baby, we therefore analyze the additional computation cost introduced by the three modules. In detail, the FS module takes $\mathcal{O}(D_{Q}\times N_Q)$ to calculate $\boldsymbol{\alpha}$, where $D_{Q}$ denotes the dimension size of $\mathbf{Q}$. Then, the FR module takes at most $\mathcal{O}(D_{Q}\times D_{R})$ to finish the feature refinement, where $D_{R}$ denotes the dimension size of the $\mathbf{Q}_{R}$. As to the FCM module, $\mathcal{O}(d_{r}^2)$ is taken to obtain the feature correlation signals. Note that the $D_Q, D_{R}$ and $d_r$ are relatively small values and $\mathcal{D}_{R}$ is much smaller than $\mathcal{D}_{Q}$, the additional computation cost is negligible. 

As to model size, on the Alimama dataset, \baby contains $15.40$ Billion~(B) parameters. In contrast, this number for some representative baselines is as follows: PLE - 15.31B; AEMS$^2$ - 15.31B; MMoE - 15.29B; STAR - 15.21B. It is clear that our model has a comparable model size but consistently better performance in different multi-scenario tasks.

\begin{table}[!]
\small
\setlength{\tabcolsep}{5pt}
\centering
\caption{The ablation study of \baby on two Datasets. The best results are highlighted in boldface.}
    \label{tab:ablation}
    \begin{tabular}{c|cccc}
    \toprule
    \multirow{2}{*}{Models}&\multicolumn{4}{c}{Alimama}\cr
    \cmidrule(lr){2-5}
    &VS&SS&IS&Total Gain\cr
    \cmidrule{1-5}
    w/o ST & 0.7461 & 0.6873 & 0.7173 & -0.0071 \cr
    w/o FCM & 0.7445 & 0.6922 & 0.7174 & -0.0037\cr
    w/o FR & 0.7466 & 0.6903 & 0.7150 & -0.0059\cr
    w/o FS & 0.7462 & 0.6909 & 0.7141  & -0.0066\cr
    w/o NL & 0.7447 & 0.6877 & 0.7169  & -0.0085\cr
    w/o GS & 0.7455 & 0.6912 & 0.7165 & -0.0046\cr
    \midrule
    \baby & \textbf{0.7473} & \textbf{0.6927} & \textbf{0.7178} & -- \cr
    \midrule
    \multirow{2}{*}{Models}&\multicolumn{4}{c}{Ali-CPP}\cr
    \cmidrule{2-5}
    &S1&S2&S3&Total Gain\cr
    \cmidrule{1-5}
    w/o ST & 0.5792 & 0.5986 & 0.5952 & -0.0330 \cr
    w/o FCM & 0.5686 & 0.5893 & 0.5877 & -0.0604\cr
    w/o FR & 0.5709 & 0.5891 & 0.5873 & -0.0587\cr
    w/o FS & 0.5844 & 0.6005 & 0.5983  & -0.0228\cr
    w/o NL & 0.5796 & 0.5989 & 0.5969  & -0.0306\cr
    w/o GS & 0.5857 & 0.6053 & 0.5997 & -0.0153\cr
    \midrule
    \baby & \textbf{0.5869} & \textbf{0.6114} & \textbf{0.6077} & --\cr
    \bottomrule
    \end{tabular}
\end{table}

\paratitle{Ablation Study.} We conduct further ablation study to validate each design choice: Feature Scaling~(FS), Feature Refinement~(FR), Feature Correlation Modeling module~(FCM), Network Layer~(NL), the shared tower (ST) in the prediction layer
 and the Gumbel Softmax~(GS) respectively. 

Table~\ref{tab:ablation} reports the performance of these variants and the full \baby model on both datasets. Here, we can make the following observations.
On the whole, these six components have played an obvious positive role in both datasets. The results illustrate that FS, FR, and FCM modules improve the ability to model complex multi-scenario situations through refinement and information sharing at the feature level. Also, the performance reduction by removing NL and ST demonstrates that we still need to facilitate knowledge transfer on top of adaptive feature learning. But we emphasize again that scenario-aware adaptive feature learning is simpler and more efficient for MSL.

\begin{figure}[!]
  \begin{subfigure}{\linewidth}
		\centerline{
            \includegraphics[width=1.1\linewidth]{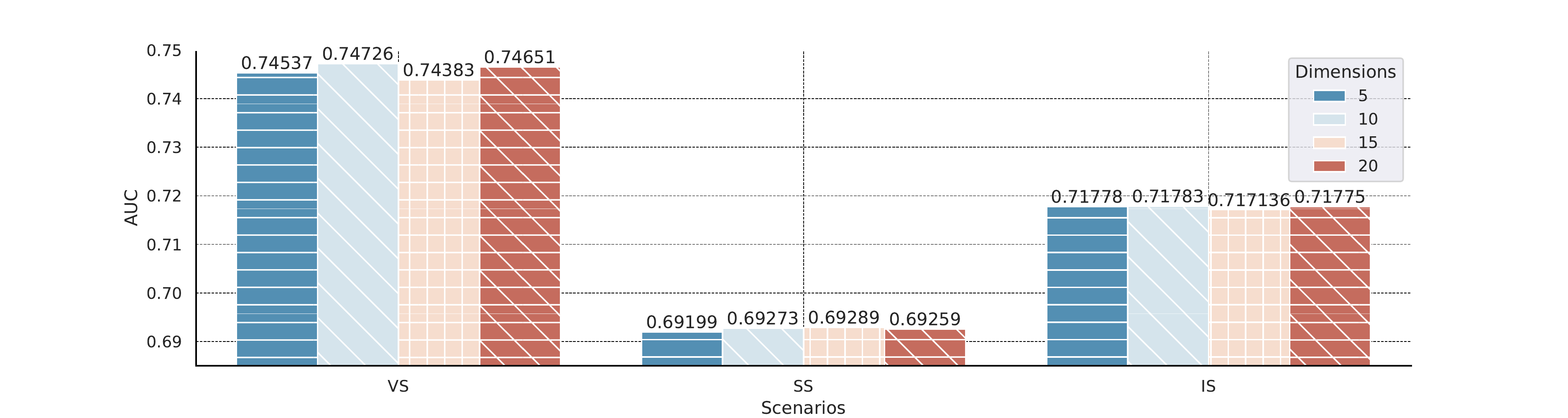}}
		\caption{On Alimama dataset}
		\label{dimension_alimama}
  \end{subfigure}
  \begin{subfigure}{\linewidth}
		\centerline{
		\includegraphics[width=1.1\linewidth]{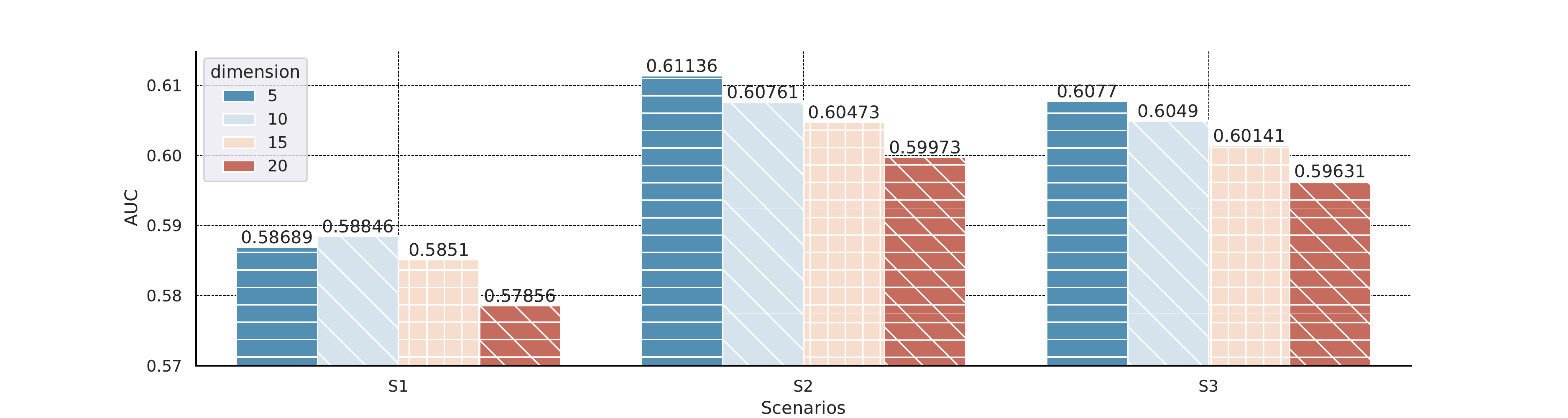}}
		\caption{On Ali-CCP dataset}
		\label{dimension_ccp}
  \end{subfigure}
  \caption{Performance for each scenario with different $d_{co}$.}
  \label{fig:dimensions}
\end{figure}

\begin{figure}[!]
  \centering
  \begin{subfigure}{0.49\linewidth}
		\centerline{
		\includegraphics[width=\linewidth, trim=0 10 0 0, clip]{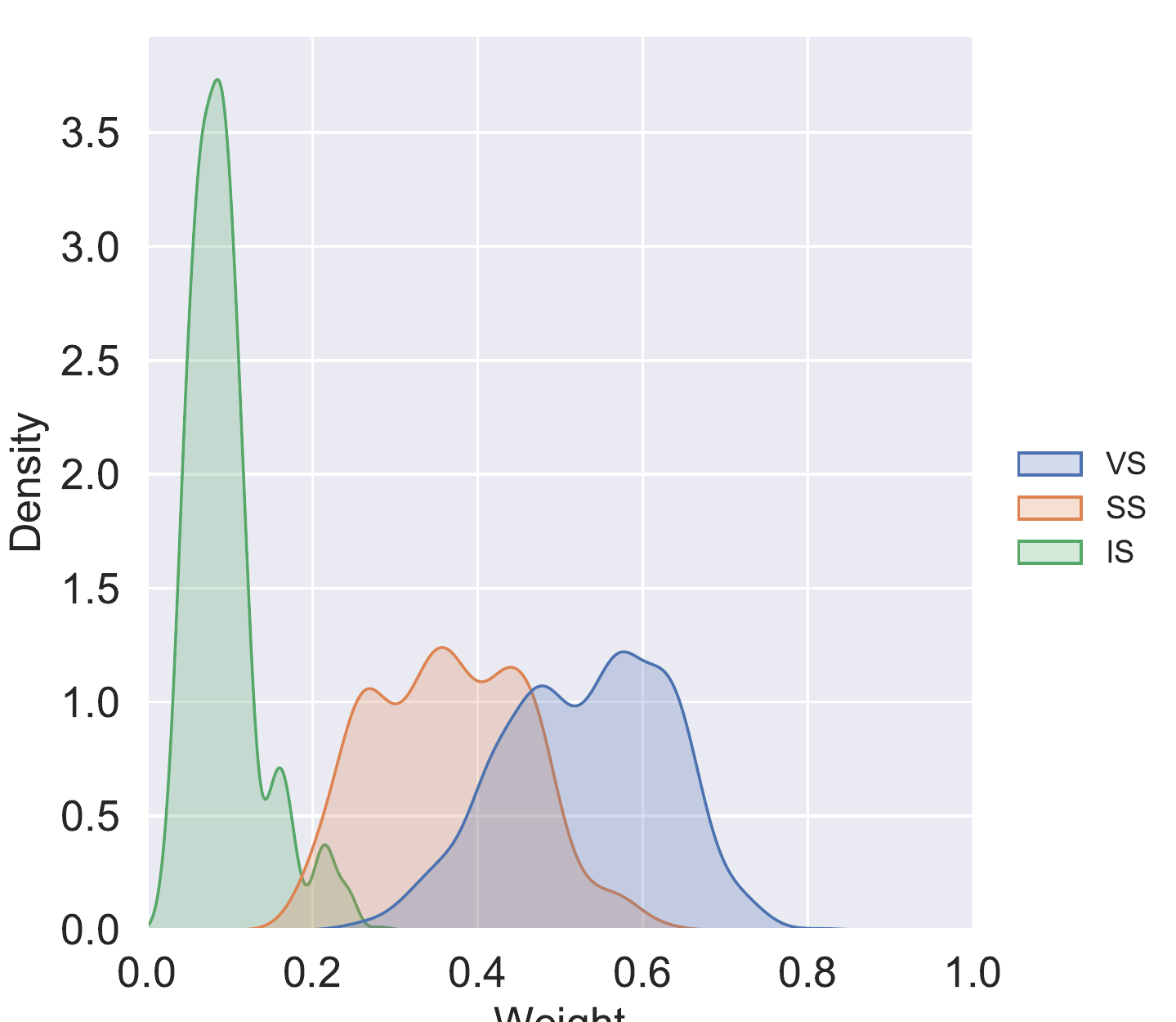}}
		\caption{User field.}
		\label{fig:FCD_sparseuser}
  \end{subfigure}
  \begin{subfigure}{0.49\linewidth}
		\centerline{
		\includegraphics[width=\linewidth, trim=0 10 0 0, clip]{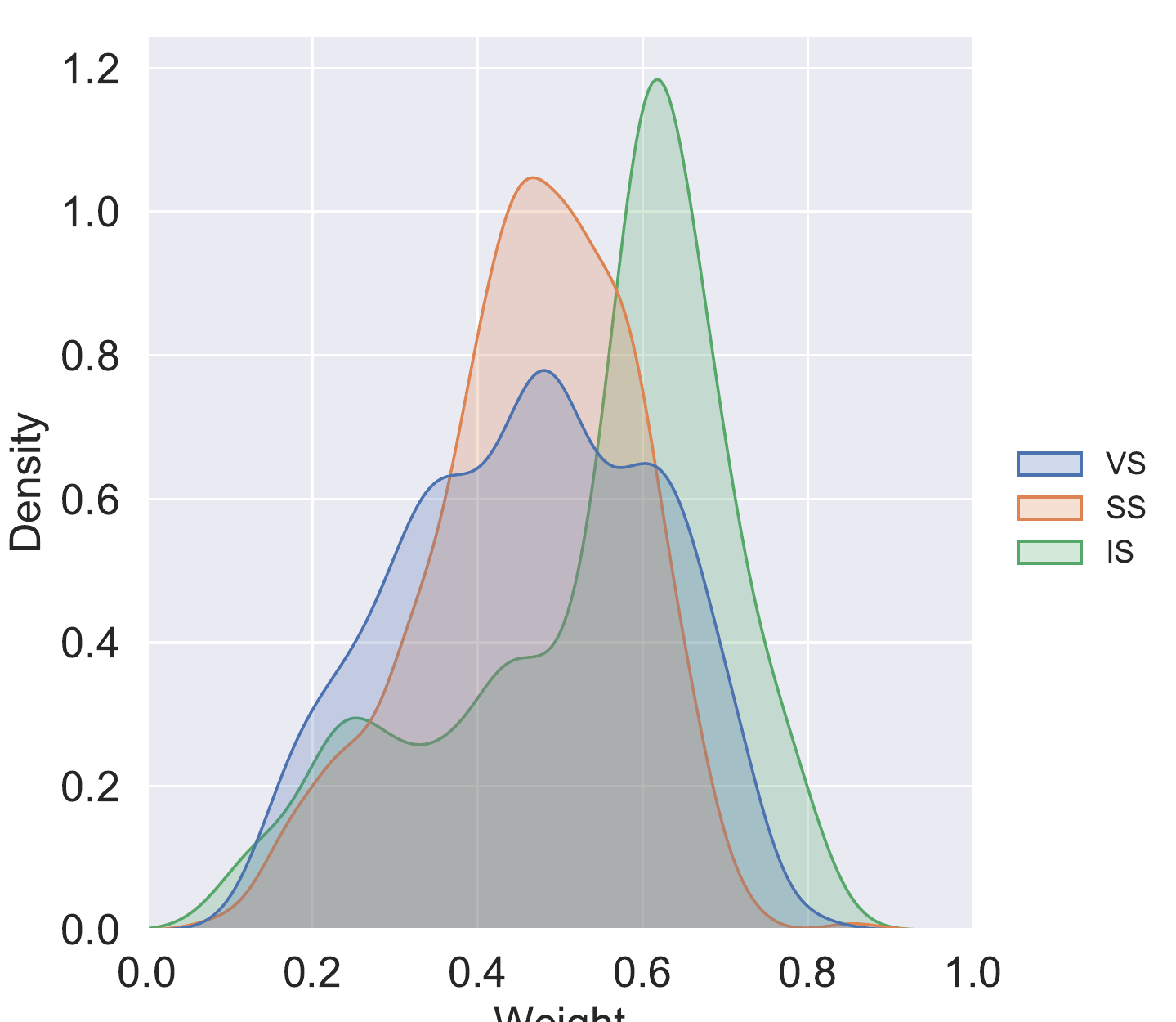}}
		\caption{User recent behavior field.}
		\label{fig:FCD_urb}
  \end{subfigure}
  \caption{Distribution of the RFC weight in the FR modules.}
  \label{fig:FR_vis}
\end{figure}

\paratitle{Impact of $d_r$ values.} Recall that we measure the semantic correlations between different feature fields by projecting them into the same dimension size $d_r$. In other words, the setting of this value controls the granularity of correlation measure. Accordingly, we plot the performance patterns of varying dimension $d_{r}$ values for both Ali-CCP and Alimama datasets in Figure~\ref{fig:dimensions}. Here, it is reasonable to see that AUC scores increase when $d_r$ becomes small. Our \baby obtains the best performance on two out of three scenarios for Ali-CCP when the dimension is $5$. Nevertheless, on the Alimama dataset, considering the tremendous data volume, more complex scenarios, and other factors, \baby is not sensitive to varying values of $d_r$ in different scenarios but a relatively large value of $10$ is more optimal.

\paratitle{Visualization of the FR module.}
A key part of \baby is the FR module for automatic refiner selection. Hence, to gain further insight into the effectiveness of this mechanism, we sample $1,000$ examples and calculate the refiner selection weights distribution on the Alimama dataset\footnote{The same phenomenon can be observed on the Ali-CCP dataset as well.}. Note that we set the number of refiners to be two for the Alimama dataset, Figure~\ref{fig:FCD_sparseuser} and Figure~\ref{fig:FCD_urb} illustrate the distribution of picking the first refiner across different scenarios on the user field and user behavior field respectively. The abscissa is the probability value, and the ordinate is the data density value obtained after the kernel function transformation. In this way, by comparing the distribution differences of the same refiner corresponding to the same feature field in different scenarios, we can prove that the adaptive feature learning mechanism has realized our expected purpose. That is, the feature-level adaptation is achieved by selecting different refiners to extract high-level semantics at an instance level. Specifically, we can see that the distribution deviation is very obvious (as shown in Figure~\ref{fig:FCD_sparseuser}). This confirms the unique characteristics of this MSL setting: more diverse scenarios with different page environments and user intents, and so on.
In this situation, the state-of-the-art solutions cannot well handle this distribution deviation using the same feature representation for all scenarios. In Figure~\ref{fig:FCD_urb}, the distribution deviation in the behavior field also significantly holds between VS and IS scenarios.

\section{Offline A/B test}
We also perform the offline A/B test based on the traffic logs of the three scenarios in Alimama. The performance comparison between \baby and three independent base models that were serving online is reported in Table~\ref{tab:offlineAB}. In real serving environments, VS, SS, and IS are trained using 60, 90, and 120 days of historical data respectively. While Maria applies 60 days of training data in all three scenarios, which greatly reduces the cost of training. We take the averaged results from March 11th to 16th, 2023, as shown in Table~\ref{tab:offlineAB}. Specifically, As for \baby, we successfully obtain $12.3\%$ of AUC revenue in total. In order to verify whether the prediction is closer to the real click-through rate~(CTR), we further measure the ratio of the Predicted CTR Over the posterior CTR~(PCOC). When PCOC is closer to $1.0$, the more accurate the CTR prediction is. In Table~\ref{tab:offlineAB}, we can see that the PCOC of \baby in all scenarios is more compact and concentrated around $1.0$ than the three serving models, which demonstrates the superiority of our solution. 
\begin{table}[!]
  \centering
  \setlength{\tabcolsep}{7pt}
  \small
  \caption{Comparisons with base serving models in all scenarios on the Alimama platform.}
  \label{tab:offlineAB}
    \begin{tabular}{ccccc}
    \toprule
    Models & Metric & VS & SS & IS \cr
    \midrule
    base models & \multirow{2}{*}{AUC} & 0.7344 & 0.5812 & 0.8279 \cr
    \baby &  & \textbf{0.7416} & \textbf{0.6832} & \textbf{0.8441} \cr
    \midrule
    base models & \multirow{2}{*}{PCOC} & 0.0737 & 1.3641 & 0.0551 \cr
    \baby &  & \textbf{0.0602} & \textbf{0.0826} & \textbf{0.0226} \cr
    \bottomrule
    \end{tabular}
\end{table}

\section{Conclusion }
In this paper, we propose a multi-scenario ranking framework with adaptive feature learning, named \baby. We design three components to enable discriminative feature learning in a scenario-aware manner: \textit{feature scaling}, \textit{feature refinement}, and \textit{feature correlation modeling}. In this way, the scenario-relevant features suppress the irrelevant counterparts. And an automatic refiner selection in the FR module further refines the high-level semantics for better scenario adaptation. Moreover, the feature correlations are also exploited to enhance discrimination. The key point of our proposed \baby is to highlight that performing adaptive feature learning could be simpler and more effective instead of conducting the structure search. 

\eat{
The rationality is that the input at the bottom layer is the fatal bottleneck: a nine-storeyed terrace must be constructed from its very base. According to the experiments on two large real-world datasets covering both product search and recommendation, our \baby obtains significantly better ranking performance than a series of baselines and SOTA alternatives.
}

\begin{acks}
    This work was supported by National Natural Science Foundation of China (No.~62272349); Alibaba Group through Alibaba Innovative Research Program; and Young Top-notch Talent Cultivation Program of Hubei Province. Chenliang Li is the corresponding author.
\end{acks}

\newpage
\balance
\bibliographystyle{ACM-Reference-Format}
\bibliography{refer}

\end{document}